\documentclass[10pt]{article}
\usepackage{amsfonts}
\usepackage{amsmath}
\usepackage{amssymb}
\usepackage{graphicx}
\usepackage{color}
\usepackage{comment}
\usepackage{cite}

\def \be {\begin{equation}}
\def \ee {\end{equation}}
\def \bea {\begin{eqnarray}}
\def \eea {\end{eqnarray}}

\def \la {\langle}
\def \ra {\rangle}
\def \rr {\raise.35ex\hbox{\small $\prime$}\kern-.17em{\mbox{\large $\imath$}}}
\def \del {\partial}
\def \dels {\partial\kern-.5em / \kern.5em}
\def \As {{A\kern-.5em / \kern.5em}}
\def \Ds {D\kern-.7em / \kern.5em}

\def \d {\delta}
\def \eps {\epsilon}

\def \lam {\lambda}

\def \s {\sigma}

\newcommand{\ba}{\begin{eqnarray}}
\newcommand{\ea}{\end{eqnarray}}

\begin{document}
\begin{titlepage}

\begin{center}

\hfill \today
\vskip .5in

\textbf{\LARGE Nambu Bracket for M Theory}

\vskip .5in
Pei-Ming Ho$^\dagger$\footnote{
E-mail address: pmho@phys.ntu.edu.tw}, 
\\
\vskip 3mm
{\it\large
$^\dagger$
Department of Physics, Center for Theoretical Sciences \\
and Leung Center for Cosmology and Particle Astrophysics, \\
National Taiwan University, Taipei 10617, Taiwan,
R.O.C.}\\

\vspace{60pt}
\end{center}
\begin{abstract}
\end{abstract}

In this article we give a concise review of recent progress 
in our understanding of the Lie 3-algebra and
their application to the Bagger-Lambert-Gustavsson model
describing multiple M2-branes in M theory.

\end{titlepage}

\setcounter{footnote}{0}

\section{Introduction}

In 1973, Nambu \cite{Nambu} proposed 
an alternative to the canonical formulation. 
In the canonical formulation, 
the equations of motion 
(Hamilton-Jacobi equations) of a physical system 
are defined
via a Poisson bracket and a Hamiltonian.
In Nambu's formulation, 
the Poisson bracket is replaced by the Nambu bracket.
The latter differs from the former by having 3 slots instead of 2.
In general, for an algebra ${\cal A}$, 
a Nambu bracket is a skew-symmetric tri-linear map 
${\cal A}\otimes {\cal A}\otimes {\cal A} \rightarrow {\cal A}$.
In Nambu's original paper \cite{Nambu},
he considered imposing various properties on 
the bracket for various applications.
It was not totally clear then, or now, 
which are the properties one should impose on the Nambu bracket
for a certain application.

Later, Takhtajan \cite{Takhtajan} defined 
a special class of Nambu brackets, called
the Nambu-Poisson brackets (See Sec. \ref{NPA}). 
He found a class of 2 dimensional (string) systems 
for which the Nambu-Poisson bracket can be used 
to write down the equations of motion 
(together with 2 Hamiltonians) with the advantage 
that a larger symmetry is kept manifest
compared with the canonical formulation, 
which requires a gauge fixing condition.

More recently, Bagger, Lambert 
\cite{Bagger:2006sk,Bagger:2007jr,Bagger:2007vi} 
and Gustavsson \cite{Gustavsson:2007vu}
proposed 
a model to describe the system of multiple M2-branes in M theory. 
The model is defined on a Lie 3-algebra, 
which is equipped with a Nambu bracket that 
satisfies a generalized Jacobi identity.
The purpose of this article is to give a concise review 
of recent advances in understanding the Lie 3-algebra 
and the BLG model.

\section{Mathematical Aspects}

\subsection{Lie 3-algebra}

Lie $n$-algebra, also known as $n$-ary Lie algebra, 
or Filippov $n$-algebra \cite{Filippov}, 
is a natural generalization of Lie algebra. 
For a linear space 
${\cal V} = \{ \sum_{a=1}^{N} v_a T_a ; v_a \in \mathbb{C} \}$
of dimension ${N}$, 
a Lie $n$-algebra structure is defined by 
a multilinear map (an $n$-bracket) 
$[ \cdot, \cdots, \cdot ]: {\cal V}^{\otimes n} \rightarrow {\cal V}$ 
satisfying the following properties 
\begin{enumerate}
\item
{\em Skew-symmetry}: 

For arbitrary $A_1, \cdots, A_n \in {\cal V}$, 
\be
[A_{\s(1)}, \cdots, A_{\s(n)}] = (-1)^{|\s|} [A_1, \cdots, A_n],
\ee
where $\s$ is a permutation of the set $(1, 2, \cdots, n)$
and $(-1)^{|\s|}$ denotes the signature of permutation.

\item
{\em Fundamental identity} (or {\em Generalized Jacobi Identity}): 

For arbitrary $A_1, \cdots, A_{n-1}, B_1, \cdots, B_n \in {\cal V}$,
\be \label{FI} 
[A_1, \cdots, A_{n-1}, [B_1, \cdots, B_n]]
= \sum_{k=1}^n [B_1, \cdots, B_{k-1}, [A_1, \cdots, A_{n-1}, B_k ], 
B_{k+1}, \cdots, B_n].
\ee
\end{enumerate}

Let us now focus on the case of Lie 3-algbera. 
There are two ways to understand the significance of 
the fundamental identity. 

First, if we define a transformation on the elements 
of a Lie 3-algebra 
\footnote{
Here $\wedge$ denotes the antisymmetrized tensor product, i.e.,
\be
A\wedge B\equiv \frac{1}{2}(A\otimes B - B\otimes A).
\ee
}
\be
\label{transfA}
\d_\Lambda A = \Lambda_{ab} [T_a, T_b, A], \qquad
\Lambda \equiv \Lambda_{ab} (T_a \wedge T_b), 
\ee
where $\Lambda_{ab}$ are the transformation parameters, 
the fundamental identity is equivalent to 
the statement that the 3-bracket (or {\em Nambu bracket})
is covariant, i.e.,
\be
\label{FI1}
\d_\Lambda [A, B, C] = [\d_\Lambda A, B, C] 
+ [A, \d_\Lambda B, C] + [A, B, \d_\Lambda C].
\ee

The transformations $\d_\Lambda$ in (\ref{transfA}) can be understood 
as generators of a Lie algebra.
The fundamental identity is also equivalent to the statement that 
the Lie bracket of the transformations is given by 
\be
\label{FI2}
[\d_{\Lambda_1}, \d_{\Lambda_2}] 
= \d_{\Lambda_3},
\ee
where 
\be
\Lambda_3 = (\Lambda_1)_{ab}(\Lambda_2)_{cd}
([T_a, T_b, T_c] \wedge T_d + T_c \wedge [T_a, T_b, T_d]).
\ee

Therefore, the 3-bracket can be used to define symmetry transformations 
in a physical model, 
and it can also be used to construct a covariant object 
from 3 covariant quantities that transforms like (\ref{transfA}).

According to the definition above, 
a usual Lie algebra is also a Lie 2-algebra. 
In physical applications, it is often convenient to realize a Lie algebra 
via a representation, which is a map
${\cal R}: {\cal V} \rightarrow M_k$,
where $M_k$ is the space of $k\times k$ matrices.
Given a representation, 
the Lie bracket 
can be realized as the commutator
\be
\label{comm}
{\cal R}([A, B]) = {\cal R}(A){\cal R}(B) - {\cal R}(B){\cal R}(A).
\ee
We also know how to construct (or decompose) representations
from (or into) irreducible representations.
For a Lie 3-algebra, a representation can be defined as a map 
from ${\cal V}\wedge{\cal V}$ to $M_k$.
It is obvious that this can be done, 
as $\delta_{\Lambda}$ defined in (\ref{transfA}) can be understood 
as a Lie algebra generator, 
and $\Lambda \in {\cal V}\wedge{\cal V}$.
However,
the drastic difference between Lie algbera and Lie 3-algebra is that 
a matrix representation for Lie algebra automatically guarantees 
the Jacobi identity to be satisfied, 
thus one can easily construct Lie algebras from the algebra of a set of matrices.
This ensures the existence of plenty of Lie algebras.
On the other hand, no representation of the Lie 3-algebra implies
the validity of the fundamental identity, 
and it is hence much harder to construct new examples of Lie 3-algebras. 
This is probably one of the greatest obstacles 
that hinder the application of Lie 3-algebra to physics.

In terms of a basis $\{T_a\}$, 
a Lie 3-algebra is defined by the structure constants $f_{abc}{}^d$:
\be
[T_a, T_b, T_c] = f_{abc}{}^d \,T_d.
\ee
The  fundamental identity implies a bilinear relation among
the structure constants,
\be
f_{b_1 b_2 b_3}{}^c f_{a_1 a_2 c}{}^{d}=
{f_{a_1 a_2 b_1}}^c {f_{c b_2 b_3}}^d +
{f_{a_1 a_2 b_2}}^c {f_{b_1 c b_3}}^d +
{f_{a_1 a_2 b_3}}^c {f_{b_1 b_2 c}}^d.
\ee

Analogous to the Killing form in Lie algebra, 
it is often necessary for physical applications that 
the Lie 3-algebra be equipped with a metric
\ba
\la T_a | T_b \ra =h_{ab}
\ea
which is invariant under the transformations (\ref{transfA}). 
That is, 
\be
\la \d_\Lambda A | B \ra + \la A | \d_\Lambda B \ra = 0.
\ee
This implies a relation for the structure constants
\ba\label{invmetric}
{f_{a_1 a_2 b}}^d h_{dc} + {f_{a_1 a_2 c}}^d h_{db}= 0.
\ea
As $f_{a_1 a_2 a_3}{}^b$ is antisymmetrized in its subscripts 
due to the skew-symmetry of the 3-bracket, 
the tensor 
\be \label{tat}
f_{a_1 a_2 a_3 a_4} \equiv f_{a_1 a_2 a_3}{}^b h_{b a_4} 
\ee
is totally antisymmetrized.

\subsection{${\cal A}_4$}
\label{A4}

The simplest nontrivial Lie 3-algebra is ${\cal A}_4$. 
It has 4 generators $\{T_a\}_{a=1}^4$. 
The Nambu bracket is defined by 
\be
[ T_a, T_b, T_c ] = \epsilon_{abcd} T_d, 
\ee
where $\epsilon_{abcd}$ is the totally antisymmetrized tensor 
with $\epsilon_{1234} = 1$. 
The invariant metric of ${\cal A}_4$ is 
\be
\la T_a | T_b \ra = \delta_{ab}.
\ee
${\cal A}_4$ is obviously a natural generalization of 
the Lie algebra $su(2)$. 

${\cal A}_4$ was named the ``generalized spin algebra''
in \cite{Kawamura:2003cw}, 
where representations of ${\cal A}_4$ in terms of matrices 
and cubic matrices \cite{Kawamura:2002yz} were constructed.
\footnote{
Cubic matrix representations of ${\cal A}_4$ were also 
given in \cite{HHM}.
}
The matrix representations in \cite{Kawamura:2003cw} 
have the peculiar feature \cite{HHM} 
that one of the generators is proportional to the identity matrix. 
There is therefore some motivation to explore 
the possibility of using cubic matrices to represent Lie 3-algebras. 
However, the cubic matrix representations do not fulfill 
the requirement of fundamental identity automatically, 
and it is not clear whether cubic matrix representations exist 
for all Lie 3-algebras. 

It was conjectured in \cite{HHM} and later proved in \cite{no-go} 
that the only finite dimensional Lie 3-algebras with 
a positive-definite metric are the trivial algebra, ${\cal A}_4$, 
and their direct sums. 
\footnote{
Similar conjectures were also proposed and studied in 
\cite{FigueroaO'Farrill:2002xg}. 
}
On the other hand, it is possible to define 
many infinite dimensional Lie 3-algebras with 
positive-definite metrics.
All the Nambu-Poisson algebras are of this kind.

\subsection{Lie 3-Aglebras with Non-Positive-Definite Metric}

As we mentioned above, finite dimensional Lie 3-algebras 
are either just copies of ${\cal A}_4$ with a trivial algebra, 
or their metrics are not positive-definite.
Thus we are forced to consider Lie 3-algebra metrics with 
non-positive-definite signatures. 

Readers who are interested in learning mathematical issues 
such as how these Lie 3-algebras are systematically constructed, 
how general or unique they are, etc. 
are directed to the literature \cite{DeMedeiros:2008zm,deMedeiros:2009hf}. 

A Lie 3-algebra with generators $\{ u_a, e_i \}$ 
($a, b, c = 1, 2, \cdots, M$, $i = 1, 2, \cdots, N$) 
can be defined by \cite{Ho:2009nk}
\bea
{}[u_a, u_b, u_c] &=& K^i_{abc} e_i + L_{abcd} v^d, \label{uuu} \\
{}[u_a, u_b, e^i] &=& J_{ab}^{ij} e_j - K^i_{abc} v^c, \label{uue} \\
{}[u_a, e^i, e^j] &=& J_{ab}^{ij} v^b + f_a^{ijk} e_k, \label{uee} \\
{}[e^i, e^j, e^k] &=& - f_a^{ijk} v^a. \label{eee}
\eea
The constant coefficients $K^i_{abc}$, $L_{abcd}$, $J_{ab}$ and $f_a^{ijk}$
are defined as follows. 

First, the range $I \equiv \{1, 2, \cdots, N\}$ of the indices $i, j, k$ 
is divided into $n$ disjoint sets $I = I_1\cup I_2 \cup \cdots \cup I_n$, 
corresponding to $n$ Lie algebras $\textbf{\em g}_a$
with structure constants $f^{ijk}_a$
and Killing form $g_a^{ij}$ 
(which is used to raise and lower indices on $e_i$ and $e^i$). 
We have 
\bea
f^{ijk}_a&=&\left\{
\begin{array}{ll}
f^{ijk}_a, \quad & i,j,k\in I_a,\\
0, & \mbox{otherwise},
\end{array}
\right. \\
J^{ij}_{ab} e_j &=&\left\{
\begin{array}{ll}
{\cal D}_{ab}(e^i), \quad & 
\mbox{both} \;\; i, j \in I_a \quad \mbox{or both} \;\; i, j \in I_b, \\
0, & \mbox{otherwise},
\end{array}
\right. \\
K_{abc}^i e_i &=&\left\{
\begin{array}{ll}
[{\cal D}_{ac}, {\cal D}_{bc}] +
[{\cal D}_{ba}, {\cal D}_{ca}] +
[{\cal D}_{cb}, {\cal D}_{ab}] +
C_{abc}, \quad & i \in I_a \cup I_b \cup I_c, \\
0, \quad & \mbox{otherwise},
\end{array}
\right.
\eea
where ${\cal D}_{ab}$ is a derivative of 
$\textbf{\em g}_a\oplus \textbf{\em g}_b$, and 
$C_{abc} = C^i_{abc} e_i$ is restricted to be a central element 
in the Lie algebras 
$\textbf{\em g}_a\oplus \textbf{\em g}_b\oplus \textbf{\em g}_c$ 
such that
\be
{\cal D}_{ab}(C_{acd}) + {\cal D}_{ac}(C_{adb}) + {\cal D}_{ad}(C_{abc}) = 0.
\ee

Let us recall that a derivative ${\cal D}$ of a Lie algebra satisfies the relation
\be
{\cal D}([e^i, e^j]) = [{\cal D}(e^i), e^j] + [e^i, {\cal D}(e^j)].
\ee
Due to the Jacobi identity, any Lie algebra element $h = h^i e_i$
defines a derivative 
\be
{\cal D}_h(e^i) \equiv [h, e^i].
\ee
They are called inner derivatives.
It turns out that if ${\cal D}_{ab}$ is an inner derivative, 
one can always carry out a change of basis on the generators $\{u_a, e^i\}$ 
such that $J_{ab}^{ij}$ vanishes for the new basis \cite{Ho:2009nk}.
Hence we only need to consider the cases when 
${\cal D}_{ab}$'s are outer derivatives. 

The nontrivial part of the invariant metric is given by 
\bea
\langle e^i | e^j\rangle &=& g_a^{ij} \qquad \mbox{if} \quad 
i, j \in I_a, \label{metric1} \\
\langle u_a | v^b \rangle &=& \delta_a^b \, , \label{metric2}
\eea
where $g_a^{ij}$ is the Killing form of the Lie algebra $\textbf{\em g}_a$. 
Even if $g^{ij}_a$ for all $a = 1, 2, \cdots, n$ are positive-definite, 
the Lie 3-algebra has $n$ negative-norm elements
$(u_a - v^a)/\sqrt{2}$.

In \cite{deMedeiros:2009hf} a slightly more general Lie 3-algebra 
was constructed. 
However, the difference does not lead to physically new configurations 
through the BLD model.

\subsubsection{Example: $M = 1$}
\label{Lorentz}

The simplest nontrivial example of the construction above 
is the following. 

For any given Lie algebra $\textbf{\em g}$ 
with generators $e^i$, 
structure constants $f^{ij}{}_k$ and Killing form $g^{ij}$, 
one can adjoint two new generators $u_0, v^0$ 
\footnote{
We used a different notation here from the above. 
Otherwise $(u_0, v^0)$ should be denoted 
$(u_1, v^1)$.
}
and construct a Lie 3-algebra defined by 
the following Lie 3-brackets \cite{GMR,Benvenuti:2008bt,HIM}
\bea
{}[u_0, e^i, e^j] &=& f^{ij}{}_k e^k, \label{uee2} \\
{}[e^i, e^j, e^k] &=& - f^{ijk} v^0, \label{eee2}
\eea
and invariant metric
\be
\la u_0 | v^0 \ra = 1, \qquad 
\la e^i | e^j \ra = g^{ij}.
\ee

This Lie 3-algebra was discovered before 
the more general algebra constructed above 
in \cite{GMR,Benvenuti:2008bt,HIM}, 
and applied to the BLG model to describe D2-branes.

\subsubsection{Example: $M > 1$}
\label{Dp-alg}

As a more complicated example, 
we start with the Lie algebra with generators 
$\{T^a_{\vec{m}}\}$ ($a = 1, 2, \cdots, N$ 
and $\vec{m} \in \mathbb{Z}^{\otimes d}$), 
\footnote{
Despite the fact that our motivation to consider 
Lie 3-algebras with non-positive-definite metrics 
was to construct finite dimensional Lie 3-algebras, 
the construction above also applies to 
infinite dimensional cases.
}
where $\vec{m}$ is a $d$-dimensional vector 
with integer components.
The Lie bracket is assumed to be of this form
\be
[T^i_{\vec{m}}, T^j_{\vec{n}}] = 
f^{ij}{}_k e^{i\theta^{ab} m_a n_b} T^k_{\vec{m} + \vec{n}}, 
\ee
where $f^{ij}{}_k$ is the structure constant of 
a Lie algebra $\textbf{\em g}$ and 
$\theta^{ab} = - \theta^{ba}$ $(a, b = 1, 2, \cdots, d)$
is a constant antisymmetric matrix.

The subscripts $\vec{m}$ on the generators $T^a_{\vec{m}}$ 
allow us to define a derivative
\be
{\cal D}_{0a}(T^i_{\vec{m}})
= m_a T^i_{\vec{m}}, \qquad 
a = 1, 2, \cdots, d.
\ee

To construct the Lie 3-algebra as described above, 
we choose the range of indices
$I = \{(i, \vec{m})\}$ to be $I = I_0$, 
that is, we choose $I_1, \cdots, I_d$ to be empty sets.
Then we have \cite{Ho:2009nk}
\bea
f^{(i\vec{l})(j\vec{m})(k\vec{n})}_0&=&
f^{ijk} e^{i\theta^{ab} l_a m_b} 
\delta^{(3)}(\vec{l} + \vec{m} + \vec{n}), \\
{\cal D}_{ab}(T^i_{\vec{m}}) &=& \left\{
\begin{array}{l}
{\cal D}_{0a}(T^i_{\vec{m}})
= m_a T^i_{\vec{m}}, \\
0, \qquad \mbox{otherwise},
\end{array}
\right. \\
K_{abc}^i e^i &=&\left\{
\begin{array}{ll}
K_{0ab} = C_{0ab}, \\
0, \quad & \mbox{otherwise},
\end{array}
\right.
\eea
with the metric
\be
\la u_a | v^b \ra = \delta_a^b, 
\qquad
\la T^i_{\vec{m}} | T^j_{\vec{n}} \ra 
= g^{ij}\delta^{(d)}(\vec{m} + \vec{n}),
\ee
where $g^{ij}$ is the Killing form for $\textbf{\em g}$.
$C_{0ab}$ are arbitrary central elements in $\textbf{\em g}$.

It turns out that the Lie 3-algebra defined this way 
is the same as the Lie 3-algebra obtained from 
the Kac-Moody algebra generated by 
$\{T^i_{\vec{m}}, D_a\}$ through the construction 
of Sec. \ref{Lorentz} \cite{Ho:2009nk}.

We can view the generators $T^i_{\vec{m}}$ as 
\be
T^i_{\vec{m}} = T^i e^{i\vec{m}\cdot\vec{x}}
\ee
where $T^i$'s are the generators 
of a Lie algebra with structure constants $f^{ij}{}_k$, 
and $e^{i\vec{m}\cdot\vec{x}}$ is the Fourier basis 
of functions on a $d$-dimensional noncommutative torus
with the noncommutativity matrix $\theta^{ab}$.
The construction of the Lie 3-algebra also applies 
to the case when functions (sections on the trivial bundle)
are replaced by sections of twisted bundles \cite{Ho:2009nk}.

\subsection{Nambu-Poisson Algebra}
\label{NPA}

A Nambu-Poisson algebra is a Lie 3-algebra of functions 
on a manifold ${\cal M}$ that satisfies, 
in addition to the skew symmetry and fundamental identity,
the Leibniz rule
\be
\{ f_1 f_2, g, h \} = f_1 \{ f_2, g, h\} + f_2 \{ f_1, g, h \} 
\ee
for arbitrary functions $f_1, f_2, g, h$ on ${\cal M}$.
Here we use the braces $\{ \cdot, \cdot, \cdot \}$ 
to denote the Lie 3-bracket, 
which is called the Nambu-Poisson bracket. 
Apparently the Nambu-Poisson structure is 
a natural generalization of the Poisson structure. 

The decomposition theorem \cite{Jacobian} states that 
locally it is always possible to choose coordinates on ${\cal M}$ 
such that 
\be
\label{NPB}
\{ f, g, h \} = \epsilon_{ijk} \del_i f \del_j g \del_k h, 
\qquad i, j, k = 1, 2, 3.
\ee
That is, only the first 3 coordinates $x^1, x^2, x^3$ 
on a local patch of ${\cal M}$ are 
relevant to the Nambu bracket. 
This theorem is analogous to, but much more restrictive than
the Darboux theorem for the Poisson algebra. 
The decomposibility theorem can be interpreted as 
the mathematical reason why there are no solitonic branes
other than 5-branes in M theory. 

The metric of the Nambu-Poisson algebra can often be defined 
by the integration on the $d$ dimensional manifold ${\cal M}$
with a suitable measure $\mu$ 
\be
\label{int-metric}
\la f | g \ra = \int d^d x \, \mu(x) f(x) g(x).
\ee

The simplest example of Nambu-Poisson algebra is
for ${\cal M} = \mathbb{R}^3$ or $T^3$ with 
the Nambu-Poisson bracket defined by (\ref{NPB}) 
with the Cartesian coordinates $x_1, x_2, x_3$.
The metric can be defined by the integration 
as in (\ref{int-metric}) with $\mu(x) = 1$.
This Nambu-Poisson algebra is an infinite dimensional 
Lie 3-algebra with a positive-definite metric.

\subsection{Lorentzian Extension of Nambu-Poisson Algebra}
\label{M5}

A Nambu-Poisson algebra can have a central extension. 
To illustrate this idea, we consider the Nambu-Poisson algebra 
on the 3D torus $T^3$.
Let us take the Fourier modes as the basis of functions
\be
\chi^{\vec{n}} = e^{2\pi i n^a x_a}, 
\ee
where $x_a$ are the Cartesian coordinates on $T^3$ 
with the equivalence relations
\be
x_a \sim x_a + m_a, \qquad m_a \in \mathbb{Z}.
\ee
The Nambu-Poisson bracket on $T^3$ can be defined as
\be
[\chi^{\vec{l}}, \chi^{\vec{m}}, \chi^{\vec{n}}] = 
(2\pi i)^3 \epsilon_{abc} l^a m^b n^c \chi^{\vec{l} + \vec{m} + \vec{n}},
\ee
with the invariant metric
\be
\la \chi^{\vec{m}} | \chi^{\vec{n}} \ra = \delta^{(3)}(\vec{m} + \vec{n}). 
\ee

Adjoining new algebraic elements $u_a$, $v^a$ $(a = 1, 2, 3)$
to the Nambu-Poisson algebra,
we define a Lie 3-algebra with the following Lie 3-brackets 
\cite{Ho:2009nk}
\bea
{}[\chi^{\vec{l}}, \chi^{\vec{m}}, \chi^{\vec{n}}] &=& 
(2\pi i)^3 \epsilon_{abc} l^a m^b n^c \chi^{\vec{l} + \vec{m} + \vec{n}}
- (2\pi i)^2 \epsilon_{abc} \delta^{(3)}(\vec{l} + \vec{m} + \vec{n}) l^a m^b v^c, \\
{}[u_a, \chi^{\vec{m}}, \chi^{\vec{n}}] &=& 
(2\pi i)^2 \epsilon_{abc} m^b n^c \chi^{\vec{m} + \vec{n}} 
- (2\pi i) \epsilon_{abc} \delta^{(3)}(\vec{m} + \vec{n}) m^b v^c, \\
{}[u_a, u_b, \chi^{\vec{n}}] &=&
(2\pi i) \epsilon_{abc} n^c \chi^{\vec{n}} - \epsilon_{abc} \delta^{(3)}(\vec{n}) v^c, \\
{}[u_a, u_b, u_c] &=& \epsilon_{abc}.
\eea
The new invariant metric has 
\be
\la \chi^{\vec{m}} | \chi^{\vec{n}} \ra = \delta^{(3)}(\vec{m} + \vec{n}), 
\qquad 
\la u_a | v^b \ra = \delta_a^b, 
\ee
with other components vanishing.

By definition, the elements $v^a$ are central, that is, 
the 3-brackets of $v^a$ with any two other elements vanish. 
If we set $v^a = 0$, 
the Lie 3-algebra is the same as the original Nambu-Poisson algebra 
with $u_a$ identified with the Cartesian coordinates $x_a$.

\section{Application to M Theory}

In M theory, the fundamental ingredients are the M2-branes 
(or membranes).
M5-branes are their solitonic electric-magnetic dual.
The action describing a single M2-brane \cite{Bergshoeff:1987cm}
and the action for a single M5-brane \cite{rM51,rM52}
are known for a long time. 
However, the action for multiple M5-branes is still unknown, 
and the action describing multiple M2-branes 
was proposed \cite{Bagger:2006sk,Bagger:2007jr,Bagger:2007vi}
only recently. 
It remains to be seen whether 
this Bagger-Lambert-Gustavsson (BLG) model of multiple membranes
will survive all the tests of validity.
In the following we will summarize supporting evidences 
for the BLG model as well as challenges to its full success.

\subsection{Bagger-Lambert Action}

Bagger and Lambert \cite{Bagger:2006sk,Bagger:2007jr,Bagger:2007vi} proposed a supersymmetric Lagrangian for a given Lie 3-algebra as 
\be
{\cal L} = -\frac{1}{2} \la D^{\mu}X^I | D_{\mu} X^I\ra 
+ \frac{i}{2} \la\bar\Psi | \Gamma^{\mu}D_{\mu}\Psi\ra 
+\frac{i}{4} \la\bar\Psi | \Gamma_{IJ} [X^I, X^J, \Psi]\ra 
-V(X) + {\cal L}_{CS},
\ee 
where $\mu = 0, 1, 2$,
$I, J, K = 1, 2, \cdots, 8$.
In the above, 
$\Psi$ is an 11D Majorana spinor satisfying the chirality condition
\be
\Gamma_{012}\Psi = - \Psi,
\ee
$D_{\mu}$ is the covariant derivative
\be
(D_\mu  \Phi)_a =
\partial _{\mu} \Phi_a -{f^{cdb}}_a A_{\mu c d}(x) \Phi_b, 
\qquad (\Phi = X^I, \Psi) 
\ee
and $V(X)$ is the potential term defined by 
\be
V(X) = \frac{1}{12}\la [X^I, X^J, X^K] | [X^I, X^J, X^K]\ra.
\ee
Finally the Chern-Simons action for the gauge potential is 
\be
{\cal L}_{CS} = \frac{1}{2}\epsilon^{\mu\nu\lam}
\left(f^{abcd}A_{\mu ab}\del_{\nu}A_{\lam cd} 
+ \frac{2}{3} f^{cda}{}_g f^{efgb} A_{\mu ab} A_{\nu cd} A_{\lam ef} \right). 
\ee

The SUSY transformation is defined by
\bea
\d X^I_a &=& i\bar{\eps}\Gamma^I \Psi_a, \\
\d \Psi_a &=& D_{\mu}X^I_a \Gamma^\mu\Gamma^I \eps  
- \frac{1}{6} X^I_b X^J_c X^K_d f^{bcd}{}_a \Gamma^{IJK}\eps, \\
\d \tilde{A}_{\mu}{}^b{}_a &=& 
i\bar{\eps}\Gamma_{\mu}\Gamma_I X^I_c \Psi_d f^{cdb}{}_a, 
\eea
where 
\be
\Gamma_{012}\epsilon = \epsilon, \qquad
\tilde{A}_{\mu}{}^b{}_a \equiv A_{\mu cd} f^{cdb}{}_a.
\ee

The gauge transformation rule is defined as
\bea
\delta X^I_a= \tilde \Lambda^b{}_a X^I_b\,,\quad
\delta \tilde A_\mu{}^b{}_a=\partial_\mu \tilde\Lambda^b{}_a-
\tilde \Lambda^b{}_c \tilde A_\mu{}^c{}_a +\tilde A_\mu{}^b{}_c
\tilde \Lambda^c{}_a\,. \label{gaugetransf}
\eea
The fermion $\Psi$ transforms like $X^I$.

While the fundamental identity is needed for the gauge symmetry 
of the multiple M2-brane theory, 
the invariant metric is also necessary to 
write down the gauge-invariant Lagrangian.

\subsection{Successes}

In the following we summarize examples of 
how different choices of Lie 3-algebras change
the physical interpretation of the Bagger-Lambert action.

\begin{itemize}

\item {\bf M2 on M-fold}

The BL action based on 
the Lie 3-algebra ${\cal A}_4$ defined in Sec. \ref{A4}
turned out to describe 2 M2-branes on an M-fold \cite{A4}.

\item {\bf M2 to D2}

The Lie 3-algebra defined in Sec. \ref{Lorentz}
was applied to the Bagger-Lambert action in 
\cite{GMR,Benvenuti:2008bt,HIM}. 
After integrating out auxiliary fields and field redefinitions, 
it was shown \cite{HIM} that the BL action precisely turns into 
the Super Yang-Mills action for D2-branes.

\item {\bf M2 to D$p$}

Plugging the Lie 3-algebra defined in Sec. \ref{Dp-alg} 
into the BL action gives the super Yang-Mills theory 
for D$p$-branes for $p = d+2$.

\item {\bf M2 to M5}

If we take the Nambu-Poisson algebra with Lorentzian extension 
in Sec. \ref{M5} as the Lie 3-algebra 
\footnote{
Actually, in \cite{HM,HIMS} we considered
the Nambu-Poisson algebra without central extension
but took a nontrivial background configuration.
It was realized in \cite{Ho:2009nk} that this is equivalent to 
taking the Lorentzian extension of Nambu-Poisson bracket 
with a trivial background.
}
in the BLG model, 
the BL action becomes a 6D theory with a self-dual 2-form gauge potential 
which should be identified with the M5-brane worldvolume theory 
in a large constant $C$-field background \cite{HM,HIMS}.
\footnote{
A hint about the connection between the Nambu-Poisson bracket 
and the M5-brane worldvolume theory in the $C$-field background 
was obtained earlier from a study of open membrane 
scattering amplitude \cite{Ho:2007vk}.
}

\end{itemize}

\subsection{Challenges}

The Bagger-Lambert action was proposed to describe 
a system of multiple M2-branes. 
This proposal suffers the following problems.

\begin{itemize}
\item
There are not so many Lie 3-algebras. 
The only nontrivial finite dimensional Lie 3-algebra 
with a positive-definite metric is ${\cal A}_4$.
The only other known example of Lie 3-algebra 
with a positive-definite metric is the Nambu-Poisson algebra.
The fundamental identity is found to be a very strong constraint 
that severely limits the existence of Lie 3-algebras. 
At the same time, we do not have a general theory 
about how to construct Lie 3-algebras, 
or enough examples for physically interesting configurations.
\item
We do not have a good intuition about the connection 
between algebraic properties of a Lie 3-algebra and 
physical degrees of freedom of M2-branes.
In contrast, we can identify open string degrees of freedom 
in the super Yang-Mills gauge theory 
describing multiple D$p$-branes.
As a result, we know how to construct D$p$-brane 
worldvolume theories living on an orbifold or orientifold.
On the other hand, the background configuration 
corresponding to a given Lie 3-algebra is often 
not clear before lengthy algebraic manipulations.
This may be related to the fact that we do not have 
a ``suitable'' representation theory for the Lie 3-algebra.
\end{itemize}

More specifically, we do not know how to construct 
Lie 3-algebras such that the BLG model describes 

\begin{itemize}
\item $N$ M2-branes in flat, uncompactified, 11D spacetime.
\item $N$ M2-branes in an M-fold other than the case 
described by ${\cal A}_4$.
\item a single M5-brane in finite $C$ field background, 
so that upon double dimension reduction it reduces 
to the noncommutative super Yang-Mills theory for a D4-brane.
\item multiple M5-branes.
\end{itemize}

These are the some of the most important topics for future study.
Whether the BLG model will become a successful model 
of multiple M2-branes depends on whether 
these problems will be solved. 
Let me conclude the paper by mentioning that 
there is a different route to describing multiple M2-branes 
via the ABJM model \cite{ABJM}. 
ABJM theory was first expressed as a theory of $N\times N$ matrices, 
but it can also be rewritten as a BLG-type theory 
with the Lie 3-algebra replaced by another type of 3-algebra 
which leads to a smaller (${\cal N} = 6$) supersymmetry \cite{N=6}.
The advantage of the ABJM model is that 
it is capable of describing an arbitrary number of M2-branes; 
its disadvantage is that part of the supersymmetry 
of the M2-brane system is not manifest. 
Another advantage of the BLG model over the ABJM model is that 
the former allows the realization of an M5-brane in a $C$-field background 
in terms of infinitely many M2-branes via the Nambu-Poisson algebra \cite{HM,HIMS}, 
while this can be done for the ABJM model only for 
particular curved (non-BPS) M5-branes \cite{Yagi}.

\subsection*{Acknowledgments}

The author thanks Chong-Sun Chu, Jos\'{e} Figueroa-O'Farrill,
Yutaka Matuso, Yosuke Imamura and Shotaro Shiba for discussions.
This work is supported in part by
the National Science Council,
and the National Center for Theoretical Sciences, Taiwan, R.O.C.



\begin{thebibliography}{99}





  
\bibitem{Nambu}
  Y.~Nambu,
  ``Generalized Hamiltonian dynamics,''
  Phys.\ Rev.\  D {\bf 7}, 2405 (1973).

\bibitem{Takhtajan}
  L.~Takhtajan,
  ``On Foundation Of The Generalized Nambu Mechanics (Second Version),''
  Commun.\ Math.\ Phys.\  {\bf 160}, 295 (1994)
  [arXiv:hep-th/9301111].
  
\bibitem{Bagger:2006sk}
  J.~Bagger and N.~Lambert,
  ``Modeling multiple M2's,''
  Phys.\ Rev.\  D {\bf 75}, 045020 (2007)
  [arXiv:hep-th/0611108].

\bibitem{Bagger:2007jr}
  J.~Bagger and N.~Lambert,
  ``Gauge Symmetry and Supersymmetry of Multiple M2-Branes,''
  Phys.\ Rev.\  D {\bf 77}, 065008 (2008)
  [arXiv:0711.0955 [hep-th]].
 
\bibitem{Bagger:2007vi}
  J.~Bagger and N.~Lambert,
  ``Comments On Multiple M2-branes,''
  JHEP {\bf 0802}, 105 (2008)
  [arXiv:0712.3738 [hep-th]].

\bibitem{Gustavsson:2007vu}
  A.~Gustavsson,
  ``Algebraic structures on parallel M2-branes,''
  arXiv:0709.1260 [hep-th].

\bibitem{Filippov}
 V. T. Filippov, ``n-Lie algebras,'' Sib. Mat. Zh.,26, No. 6, 126Ð140 (1985).

\bibitem{Kasymove}
Sh. M. Kasymov, ``Theory of n-lie algebras'',
Algebra and Logic {\bf 26}, No.3, 155-166 (1987),
translated from Algebra i Logika, {\bf 26}, No. 3, 277Ð297 (1987).

\bibitem{Kawamura:2003cw}
  Y.~Kawamura,
  ``Cubic matrix, generalized spin algebra and uncertainty relation,''
  Prog.\ Theor.\ Phys.\  {\bf 110}, 579 (2003)
  [arXiv:hep-th/0304149].

\bibitem{Kawamura:2002yz}
  Y.~Kawamura,
  ``Cubic matrix, Nambu mechanics and beyond,''
  Prog.\ Theor.\ Phys.\  {\bf 109}, 153 (2003)
  [arXiv:hep-th/0207054].

\bibitem{HHM}
P.-M. Ho, R.-C. Hou and Y. Matsuo,
``Lie 3-Algebra and Multiple M2-branes,''
JHEP {\bf 0806} (2008) 020,
[arXiv:0804.2110 [hep-th]]. 

\bibitem{no-go}
  J.~P.~Gauntlett and J.~B.~Gutowski,
  ``Constraining Maximally Supersymmetric Membrane Actions,''
  arXiv:0804.3078 [hep-th]; 
  G.~Papadopoulos,
  ``M2-branes, 3-Lie Algebras and Plucker relations,''
  JHEP {\bf 0805}, 054 (2008)
  [arXiv:0804.2662 [hep-th]].
  
\bibitem{FigueroaO'Farrill:2002xg}
  J.~M.~Figueroa-O'Farrill and G.~Papadopoulos,
  ``Pluecker-type relations for orthogonal planes,''
  arXiv:math/0211170.

\bibitem{DeMedeiros:2008zm}
  P.~De Medeiros, J.~M.~Figueroa-O'Farrill and E.~Mendez-Escobar,
  ``Lorentzian Lie 3-algebras and their Bagger-Lambert moduli space,''
  JHEP {\bf 0807}, 111 (2008)
  [arXiv:0805.4363 [hep-th]]; 
  J.~M.~Figueroa-O'Farrill,
  ``Lorentzian Lie n-algebras,''
  J.\ Math.\ Phys.\  {\bf 49}, 113509 (2008)
  [arXiv:0805.4760 [math.RT]]; 
  P.~de Medeiros, J.~M.~Figueroa-O'Farrill and E.~Mendez-Escobar,
  ``Metric Lie 3-algebras in Bagger-Lambert theory,''
  JHEP {\bf 0808}, 045 (2008)
  [arXiv:0806.3242 [hep-th]]; 
  J.~M.~Figueroa-O'Farrill,
  ``Metric Lie n-algebras and double extensions,''
  arXiv:0806.3534 [math.RT]; 
  P.~de Medeiros, J.~Figueroa-O'Farrill, E.~Mendez-Escobar and P.~Ritter,
  ``On the Lie-algebraic origin of metric 3-algebras,''
  Commun.\ Math.\ Phys.\  {\bf 290}, 871 (2009)
  [arXiv:0809.1086 [hep-th]]; 
  J.~M.~Figueroa-O'Farrill,
  ``Three lectures on 3-algebras,''
  arXiv:0812.2865 [hep-th].

\bibitem{deMedeiros:2009hf}
  P.~de Medeiros, J.~Figueroa-O'Farrill, E.~Mendez-Escobar and P.~Ritter,
  ``Metric 3-Lie algebras for unitary Bagger-Lambert theories,''
  JHEP {\bf 0904}, 037 (2009)
  [arXiv:0902.4674 [hep-th]].

\bibitem{Ho:2009nk}
  P.~M.~Ho, Y.~Matsuo and S.~Shiba,
  ``Lorentzian Lie (3-)algebra and toroidal compactification of M/string
  theory,''
  JHEP {\bf 0903}, 045 (2009)
  [arXiv:0901.2003 [hep-th]].

\bibitem{A4}
  N.~Lambert and D.~Tong,
  ``Membranes on an Orbifold,''
  arXiv:0804.1114 [hep-th]; 
  J.~Distler, S.~Mukhi, C.~Papageorgakis and M.~Van Raamsdonk,
  ``M2-branes on M-folds,''
  arXiv:0804.1256 [hep-th].

\bibitem{GMR}
J. Gomis, G. Milanesi and J. G. Russo,
``Bagger-Lambert Theory for General Lie Algebras,''
JHEP {\bf 0806} (2008) 075,
[arXiv:0805.1012 [hep-th]].

\bibitem{Benvenuti:2008bt}
  S.~Benvenuti, D.~Rodriguez-Gomez, E.~Tonni and H.~Verlinde,
  ``N=8 superconformal gauge theories and M2 branes,''
  arXiv:0805.1087 [hep-th].

\bibitem{HIM}
P.-M. Ho, Y. Imamura and Y. Matsuo,
``M2 to D2 revisited",
arXiv:0805.1202 [hep-th].

\bibitem{Jacobian}
R. Weitzenb\"{o}k, 
``Invariantentheorie,'' P. Noordhoff, Gr\"{o}ningen, 1923.
Ph. Gautheron, 
``Some remarks concerning Nambu mechanics,''
Lett. in Math. Phys. {\bf 37} (1996), 103.
D. Alekseevsky, P. Guha, 
``On Decomposability of Nambu-Poisson Tensor,''
Acta. Math. Univ. Commenianae {\bf 65} (1996), 1.
R. Ib\'{a}nez, M. de Le\'{o}n, J. C. Marrero, D. M. de Diego, 
``Dynamics of generalized Poisson and Nambu-Poisson brackets,''
J. of Math. Physics {\bf 38} (1997), 2332.
N. Nakanishi, 
``On Nambu-Poisson Manifolds,''
Reviews in Mathematical Physics {\bf 10} (1998), 499.
G. Marmo, G. Vilasi, A. M. Vinogradov, 
``The local structure of n-Poisson and n-Jacobi manifolds,''
J. Geom. Physics {\bf 25} (1998), 141.

\bibitem{Bergshoeff:1987cm}
  E.~Bergshoeff, E.~Sezgin and P.~K.~Townsend,
  ``Supermembranes and eleven-dimensional supergravity,''
  Phys.\ Lett.\  B {\bf 189}, 75 (1987).

\bibitem{rM51}
  P.~Pasti, D.~P.~Sorokin and M.~Tonin,
  ``Covariant action for a D = 11 five-brane with the chiral field,''
  Phys.\ Lett.\  B {\bf 398}, 41 (1997)
  [arXiv:hep-th/9701037];
    I.~A.~Bandos, K.~Lechner, A.~Nurmagambetov, 
    P.~Pasti, D.~P.~Sorokin and M.~Tonin,
  ``Covariant action for the super-five-brane of M-theory,''
  Phys.\ Rev.\ Lett.\  {\bf 78}, 4332 (1997)
  [arXiv:hep-th/9701149].

\bibitem{rM52}
    M.~Aganagic, J.~Park, C.~Popescu and J.~H.~Schwarz,
  ``World-volume action of the M-theory five-brane,''
  Nucl.\ Phys.\  B {\bf 496}, 191 (1997)
  [arXiv:hep-th/9701166].

\bibitem{HM}
P.-M. Ho and Y. Matsuo,
``M5 from M2,''
JHEP {\bf 0806} (2008) 105,
[arXiv:0804.3629 [hep-th]].

\bibitem{HIMS}
P.-M. Ho, Y. Imamura, Y. Matsuo and S. Shiba,
``M5-brane in three-form flux and multiple M2-brane,''
JHEP {\bf 0808} (2008) 014,
[arXiv:0805.2898 [hep-th]].

\bibitem{Ho:2007vk}
  P.~M.~Ho and Y.~Matsuo,
  ``A toy model of open membrane field theory in constant 3-form flux,''
  Gen.\ Rel.\ Grav.\  {\bf 39}, 913 (2007)
  [arXiv:hep-th/0701130].

\bibitem{ABJM}
  O.~Aharony, O.~Bergman, D.~L.~Jafferis and J.~Maldacena,
  ``N=6 superconformal Chern-Simons-matter theories, M2-branes and their
  gravity duals,''
  JHEP {\bf 0810}, 091 (2008)
  [arXiv:0806.1218 [hep-th]].

\bibitem{N=6}
  J.~Bagger and N.~Lambert,
  ``Three-Algebras and N=6 Chern-Simons Gauge Theories,''
  Phys.\ Rev.\  D {\bf 79}, 025002 (2009)
  [arXiv:0807.0163 [hep-th]].

\bibitem{Yagi}
  S.~Terashima and F.~Yagi,
  ``M5-brane Solution in ABJM Theory and Three-algebra,''
  arXiv:0909.3101 [hep-th].



\end{thebibliography}
\end{document}